\documentclass[aps,prl,preprint,superscriptaddress]{revtex4-2}

\usepackage{physics}
\usepackage{graphicx}
\usepackage{bm}

\begin{document}

\title{Controlled transport of stored light}

\author{Wei Li}
\affiliation{School of Instrumentation and Optoelectronic Engineering, Beihang University, 100191 Beijing, China}
\affiliation{Institut f{\"u}r Physik, Johannes Gutenberg-Universit{\"a}t Mainz, 55122 Mainz, Germany}

\author{Parvez Islam}
\author{Patrick Windpassinger}
\email{windpass@uni-mainz.de}
\affiliation{Institut f{\"u}r Physik, Johannes Gutenberg-Universit{\"a}t Mainz, 55122 Mainz, Germany}

\date{\today}

\begin{abstract}
Controlled manipulation, storage and retrieval of quantum information is essential for quantum communication and computing. Quantum memories for light, realized with cold atomic samples as the storage medium, are prominent for their high storage efficiencies and lifetime. We demonstrate the controlled transport of stored light over $1.2~\textrm{mm}$ in such a storage system and show that the transport process and its dynamics only have a minor effect on the coherence of the storage. Extending the presented concept to longer transport distances and augmenting the number of storage sections will allow for the development of novel quantum devices such as optical race track memories or optical quantum registers.

\end{abstract}

\maketitle

Optical quantum memories, which allow for the storage and on demand retrieval of quantum information carried by light, are essential for scalable quantum communication networks, for instance as important building blocks in quantum repeaters \cite{Briegel1998, duan2001m} or as tools in linear quantum computing \cite{Knill2001}. A very well studied realization of light storage is based on so-called electromagnetically induced transparency (EIT) \cite{Fleischhauer_2005, Dudin_2013}. 
Here, incident light pulses are trapped and coherently mapped into collective excitation of the storage medium, which forms strongly coupled light-matter quasi-particles, the so-called dark-state polaritons (DSPs) \cite{Fleischhauer_2000}. Employing a control beam, the transparency of the medium can be switched on and off and thus light stored and retrieved from the medium. EIT-based light storage has been presented by various platforms with different application foci., which prove its high efficiency \cite{Hsiao_2018}, long lifetime\cite{Dudin_2013, Katz_2018, heinze2013stopped}, and capable of storing "true" quantum states of light \cite{eisaman2005electromagnetically, Appel2008, Vernaz-Gris2018, Nicolas2014}.

A particularly well-studied case of the storage medium is cold neutral atoms in optical dipole traps, including optical lattices \cite{Dudin_2013, yang2016efficient, Schnorrberger_2009} and optical evanescent field \cite{Corzo2019} as such systems allow for a particularly high degree of control over the parameters of the medium. The benchmark parameter "storage efficiency" in general depends on the coupling strength of light and matter \cite{Gorshkov_2007_Free}. For an ensemble of atoms, the coupling strength can be quantified by the optical depth (OD) \cite{Bajcsy_2009}. Strong coupling usually requires a tight focusing of the light beams, which in turn results in short coupling distances due to the limited Rayleigh range. This can be circumvented by trapping longitudinally extended atomic samples in or close to a quasi-one-dimensional waveguide, such as tapered optical nano-fibers \cite{Sayrin:15}, hollow-core photonic crystal fibers (HC-PCF) \cite{Blatt_2016,Bajcsy_2011,sprague2014broadband}. With the resulting extreme OD, for instance, a storage efficiency of  $23\%$ at a storage time of $0.6~\mu \textrm{s}$ has been achieved with $2\times10^5$ atoms inside an HC-PCF  system \cite{Blatt_2016}.

Apart from efficient, long-lived storage, the transport of quantum information between spatially separated locations is a crucial asset for quantum communication networks and distributed quantum computation. For instance, it is common to shuttle atoms around in ion-based quantum computer systems for various gate operations \cite{Kaufmann2017, Vogel2019} or to entangle atoms over long distances \cite{Lettner2011, Ritter2012}. By being mapped into the spin coherence of cold atoms, microwave photons have been transported in a controlled way over macroscopic distances with optical conveyor belts in free space \cite{ Kuhr2003} or though HC-PCFs \cite{Xin2019}. Compared to transporting microwave photons, the transport of stored light, or collective excitation, could have advantages due to the so-called collective enhancement, such as strong light-matter coupling strength and robustness against particle loss \cite{Lukin_2003}. However, the transport of stored light has only been demonstrated in diffusive or free-flying mediums \cite{Zibrov_2002, Xiao_2008, Ginsberg2007}.

\begin{figure*}
    \includegraphics{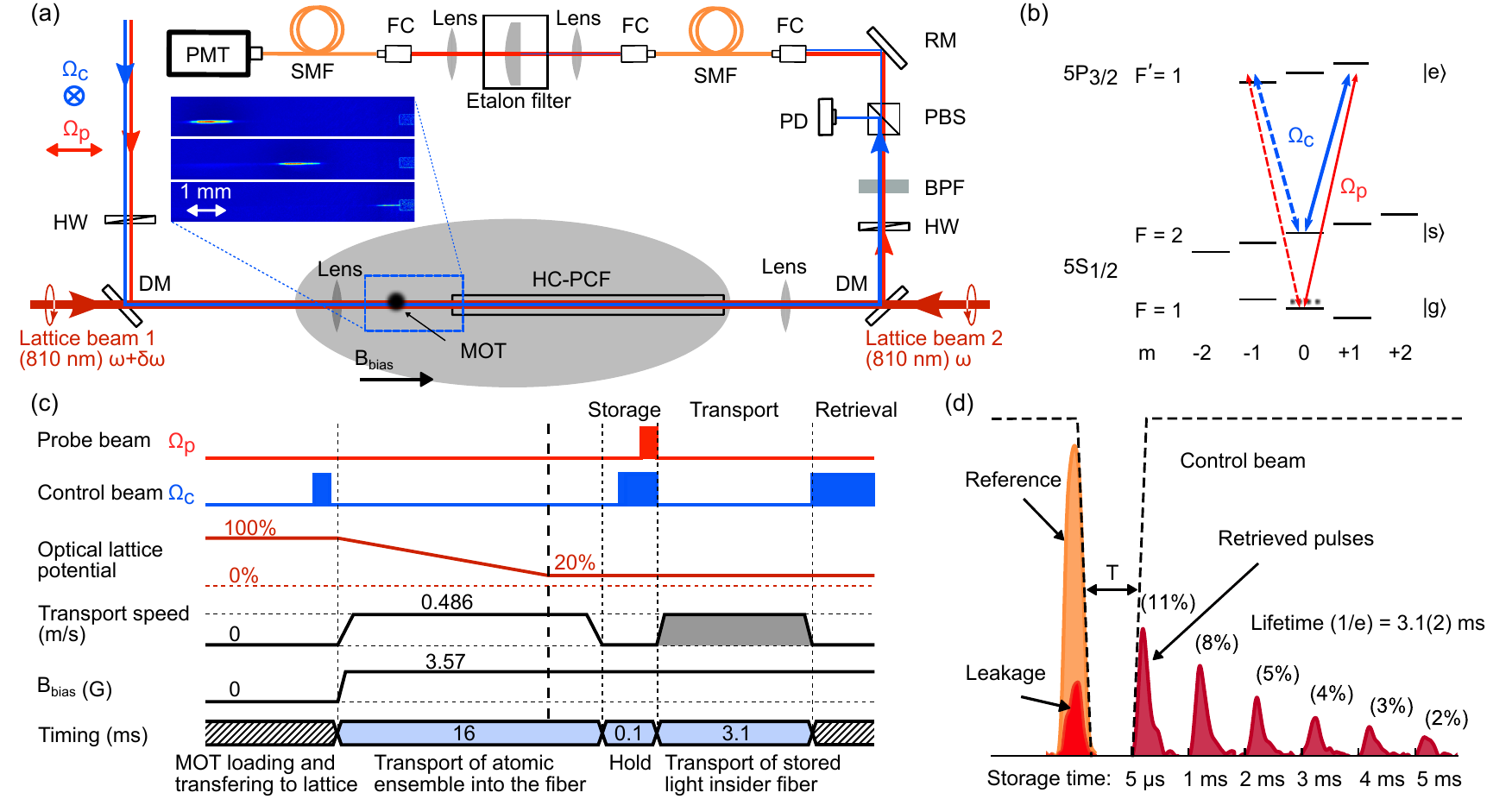}
    \caption{\label{fig:Setup}Experimental details.
        (a) Experimental setup. HW: half waveplate. DM: dichroic mirror. BPF: band-pass filter. PBS: polarizing beam splitter. PD: photodiode. RM: reflection mirror. FC: fiber coupler. SMF: single-mode fiber. PMT: photomultiplier tube. A 10-cm-long HC-PCF with a core diameter of $60~\mu \textrm{m}$ and mode field diameter of $42~ \mu \textrm{m}$ is installed inside the vacuum chamber \cite{Benabid2011}. The 3-D MOT is located $6.3~ \textrm{mm}$ in front of the fiber tip. The inset panel shows typical absorption images of the trapped atoms at the MOT position, transported by 3.3 mm and 6.3 mm, respectively.
        (b) Energy levels of the  $^{87}\textrm{Rb} ~D2$ line relevant for the experiment.
        (c) Typical experimental protocol for transport and storage. The thick vertical dashed line indicates the time when the cold atoms reach the fiber tip.
        (d) Light storage and retrieval inside the HC-PCF. T: storage time. The reference and leakage pulses are scaled by $0.5$. The x-axis is not to scale. For simplicity, the retrieved pulses of independent experimental runs with different storage time are plotted in the same figure, and only the control beam (depicted as dashed lines, not to scale) for the storage time $T=5~\mu \textrm{s}$ is shown. Each of the pulses are averaged results of eight experimental runs.}
\end{figure*}

Here, we report on actively controlled transport of stored light over macroscopic distances, i.e. distances larger than the size of the storage medium. To this end, we store light in an ensemble of cold atoms inside an HC-PCF using an EIT process. The whole ensemble containing the DSPs is then transported with an optical conveyor belt by several millimeters and the light pulse is retrieved. We benchmark our system and discuss the limitations of the transport process. By extending the experimental protocol in the future, a race track memory for light with different reading and writing sections \cite{Juzeliunas2003} comes within reach.

The experimental detail is sketched in Fig. \ref{fig:Setup}(a). We laser cool $^{87}\textrm{Rb}$ atoms in a magneto-optical trap (MOT) and transfer them into a red-detuned optical lattice. The latter is formed by two counter-propagating circularly polarized beams at $ 810~ \textrm{nm}$. The trapped atomic ensemble, with $1/e$ width of $ 1.2~ \textrm{mm}$ along the lattice axis, can be transported into the HC-PCF by detuning the frequency of the lattice beams with respect to each other \cite{Okaba2014, Langbecker2018}. To compensate for the differential ac-Stark shift caused by the inhomogeneous trapping potential, we apply a bias magnetic field ($ B_{bias}$) of $ 3.57~\textrm{G}$ along the fiber axis, which also defines the quantization axis of the system \cite{Derevianko_2010}. For the EIT storage protocol, we employ the  $\ket{g}=\ket{5S_{1/2}, ~F=1}\rightarrow \ket{e}=\ket{5P_{3/2},~ F'= 1}$ transition for the light storage, with the respective beam commonly termed probe beam ($\Omega_p$), and the $\ket{s}=\ket{5S_{1/2},~ F=2} \rightarrow \ket{e}$ transition for the control beam ($\Omega_c$), compare Fig. \ref{fig:Setup}(b). The probe and control beams are overlapped with the lattice beams at a dichroic mirror and all beams are carefully coupled to the fundamental mode of the HC-PCF with coupling efficiencies larger than 88\% \cite{Supplemental_material}.

Before the transport sequence [see Fig. \ref{fig:Setup}(c)], we optically pump the atoms to state $\ket{g}$ to reduce atom loss due to hyperfine-changing collisions in the lattice. About $ 1.2\times 10^5$ atoms are transported into the HC-PCF for storage experiments. Inside the fiber, the atomic ensemble with a radial temperature of $ 190~\mu \textrm{K}$ is trapped by the optical lattice with a trapping depth of  $ 740~\mu \textrm{K}$, corresponding to trapping frequencies of $ \omega_z = 2\pi \times 460~ \textrm{kHz}, \omega_r = 2\pi \times 4~ \textrm{kHz}$ in the axial and radial directions, respectively. The effective OD inside the fiber is measured to be $ 5$. At this moderate OD, the effects of micro-lensing can be neglected \cite{Noaman2018}. After the transport, we prepare the atoms in the $\ket{5S_{1/2},~F=1,~m=0}$ Zeeman state. This reduces photon-loss of the probe beam that is otherwise caused by the off-resonant excitation of the other Zeeman levels \cite{Vernaz-Gris2018}. Besides, using the  $\textit{m}=0$ states for storage reduces the decoherence due to magnetic field noise. The two main EIT-channels are represented by solid and dash lines as shown in Fig. \ref{fig:Setup}(b). Once the atoms have reached the desired position for storage and are prepared in the appropriate state, we switch on the control beam with a power of $ 2.7~\mu \textrm{W}$ ($ \Omega_c = 1.4~\Gamma$ with $ \Gamma$ the nature linewidth) and send a probe pulse with a full-width at half-maximum of $ 0.4 ~ \mu \textrm{s}$ and a peak power of $ 16~\textrm{nW}$. Subsequently, the probe pulse is dramatically slowed down and spatially compressed into the atomic ensemble. By reducing the power of the control beam to zero, the probe pulse is mapped onto the long-lived collective excitation of the atoms, the DSP. To retrieve the light after a variable storage time T, the control field is turned on again. The different light fields that pass through the HC-PCF are separated in a series of filters and the probe field is detected with a photomultiplier tube \cite{Supplemental_material}.

To benchmark our system, we first characterize light storage and retrieval at a fixed position of $ 1~\textrm{mm}$ from the tip, inside the HC-PCF, as shown in Fig. \ref{fig:Setup}(d). We define the storage efficiency as the ratio between the integrated energy of the retrieved pulse, and a reference pulse when there are no atoms present. The storage lifetime $ \tau$ is determined by fitting an exponential function $\textrm{exp}(-t/{\tau})$ to the data. We observe a maximum storage efficiency of $ 11(1)\%$ ($T =  5~\mu \textrm{s}$) which is comparable to other fiber-based systems \cite{Sayrin:15, Blatt_2016}. The observed lifetime of $\tau =  3.1(2)~ \textrm{ms}$ is three orders of magnitude longer than the ones reported in other HC-PCF based systems \cite{Blatt_2016, sprague2014broadband}. In our current setup, the lifetime is mainly limited by the off-resonant light scattering events from the broad background spectral of the lattice beams \cite{Supplemental_material}. As the observed storage lifetimes are larger than the time we typically need to transport the atomic sample over a distance that is larger than the sample size itself, they are sufficient for a proof-of-concept of transporting stored light.

\begin{figure}
    \includegraphics{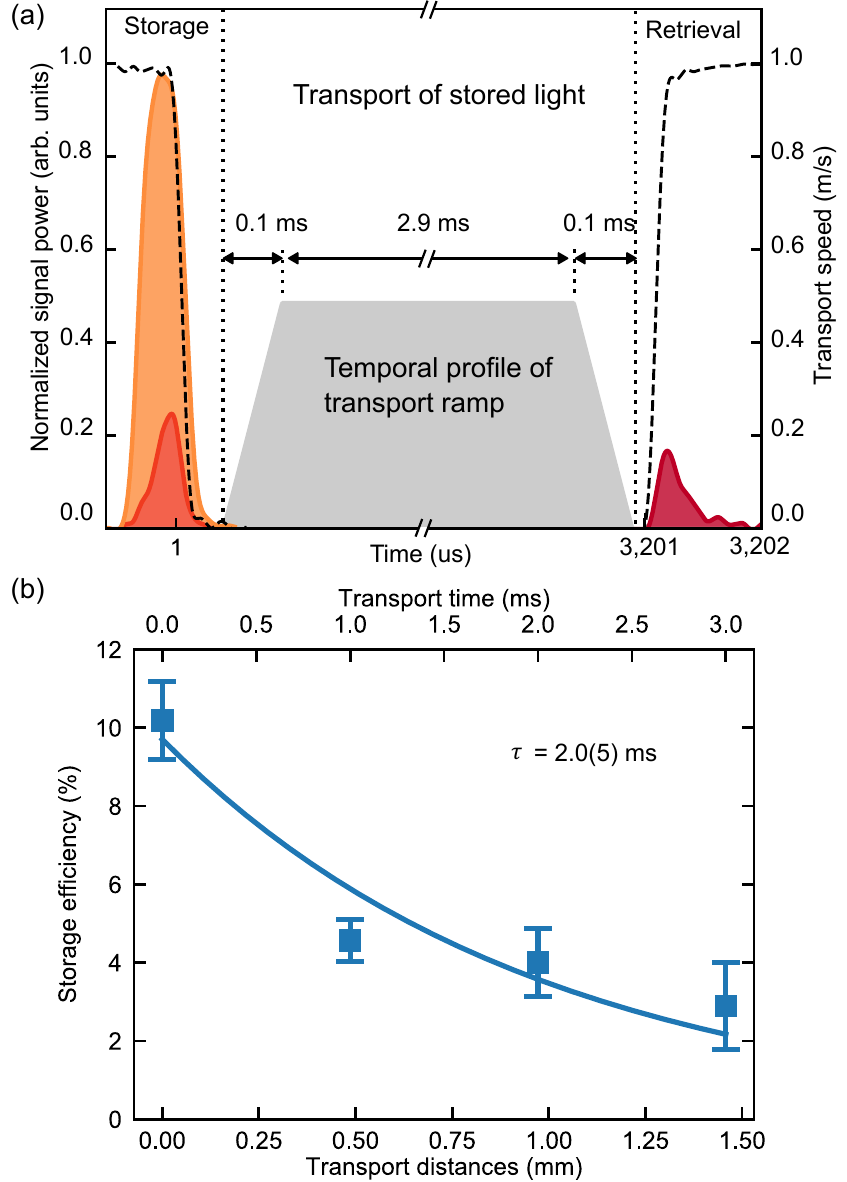}
    \caption{\label{fig:Transport_inside_fiber}
        Transport of stored light inside HC-PCF.
        (a) Typical protocol for transporting stored light with transporting time of $3.1~\textrm{ms}$. The yellow and red shadows to the left represent the reference and leakage pulse scaled by 0.25, respectively. The gray shaded area indicates the transport of the stored light and the transport speed is shown on the right axis. In red to the right, the retrieved light pulse after being transported. The control beam is depicted as the dashed lines (not to scale).
        (b) Storage efficiency for retrieval after different transport distances. Each data point is an average of eight independent runs of the experiment. Error bars represent statistical errors, which are ±1 standard deviation.. The solid line shows the fit results of the exponential function $\textrm{exp}(-t/{\tau})$.}
\end{figure}

\begin{figure*}
    \includegraphics{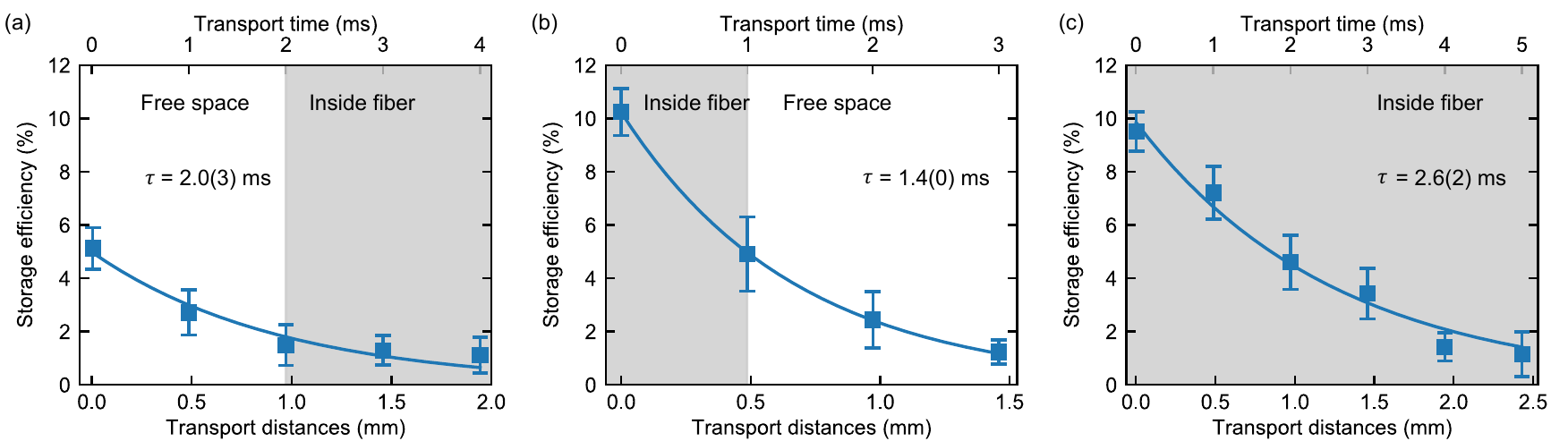}
    \caption{\label{fig:Transport_comparison}
        Transport of stored light through the interface between free space and HC-PCF.
        (a) Storage efficiency for transporting stored light from free-space into the HC-PCF. The grey shaded area indicates the fiber position.
        (b) Storage efficiency for transporting stored light out of HC-PCF.
        (c) Storage efficiency for light in a co-moving atomic ensemble inside HC-PCF. Each data point is an average of eight independent runs of the experiment. Error bars represent statistical errors, which are ±1 standard deviation.. The solid line shows the fit results of the exponential function $\textrm{exp}(-t/{\tau})$.}
\end{figure*}

The central aspect of the manuscript is that we can now actively transport these stored light pulses by switching the optical conveyor belt on again. The experimental sequence [compare Fig. \ref{fig:Transport_inside_fiber}(a)] continues as follows: After the atoms have been transported into the fiber and the transport has stopped, the probe pulse is stored into the stationary cloud of atoms. We then increase the frequency difference between the lattice beams linearly in $ 0.1~\textrm{ms}$ to accelerate the atoms, transport them for a variable time, i.e. a distance up to several millimeters and stop the transport again in $ 0.1~\textrm{ms}$. 
The acceleration is kept smaller than the maximum acceleration $ a_{\textrm{max}}=U_0 k_L / m = 5.5\times 10^5 ~\textrm{m/s}^2$ possible to reduce the loss of atoms. Here $ U_0$ is the trap depth and $ k_L$ is the wavenumber of the lattice beams. For the proof of principle discussed here, we use transport times of up to $ 3~\textrm{ms}$ which corresponds to distances ranging from $ 0.49$ to $ 1.46~\textrm{mm}$. From the data presented in Fig. \ref{fig:Transport_inside_fiber}(b), which shows the retrieval efficiency after transport, we extract that the transport does only moderately decrease the storage efficiency and is mainly limited by the storage lifetime. For instance, after $ 3~\textrm{ms}$ transport time, which corresponds to a transport distance already longer than the sample size, we obtain a storage efficiency of $ 3\%$, while the stationary cloud shows $ 4\%$ efficiency. This demonstrates that the abrupt acceleration of the sample, which for instance results in a temperature increase of the cloud \cite{Langbecker2018}, has only a moderate effect on the coherence of the DSPs. Other possible influences from the fiber itself, like van-der-Waals forces are negligible in our case due to the relatively large ratio between hollow-core size and radial diameter ($ 1/e$) of the atomic ensemble (about $ 16~ \mu \textrm{m} $ calculated from the radial temperature).
Although it is certainly also possible to achieve efficient transport of stored light in other, e.g. free space settings, the hollow-core fiber approach offers several advantages. As all the relevant light fields are guided in the fundamental mode of the fiber, they are naturally mode matched, and thus the atomic cloud confined by the lattice beams is completely embedded in the EIT beams and all the atoms participate in the storage process. This ensures the stability and high efficiency of the light-matter interaction during the whole process, which, for instance, significantly suppresses diffusion induced dephasing of the spin-wave which is observed when the individual atoms move in and out of the interaction region \cite{Dudin_2013}. Besides, thanks to the collinear configuration of the EIT beams, the spin-wave has a maximized period of $1/|\Delta\vec{k}|\approx44~\textrm{mm}$, which is orders longer than the lattice period and the thermal motion induced dephasing can be neglected.

To further substantiate our claim that we, indeed, transport the stored light in the form of collective excitation of atomic medium, we demonstrate the transport from free space to inside the fiber and in the opposite direction. This has the clear advantage that we can use standard absorption imaging techniques outside the fiber to image the position of the atomic ensemble. Sample pictures of the atom transport are included in Fig. \ref{fig:Setup}(a). The resulting data is shown in Fig. \ref{fig:Transport_comparison}(a) and (b). The individual data in each of the figures indicates the total loss of stored light due to the limited lifetime and the transport process. In the first case, we store the light $ 1~\textrm{mm}$ outside the fiber and determine the storage efficiency after moving the ensemble containing the stored light to different positions close to and into the fiber. The observed overall storage efficiency is only half of that in the case of all-in-fiber transport, compare Fig. \ref{fig:Transport_inside_fiber}(b). However, the effective storage lifetime is similar. We attribute this lower efficiency to the weaker light-matter interaction outside the fiber, where the atomic sample is less dense due to the natural divergence of the light fields.
Similar effects are expected when moving the stored light in the opposite direction. To demonstrate this, we first bring the atomic sample to a position of $ 0.5~\textrm{mm}$ inside the fiber, store the light pulse and then reverse the transport direction of the conveyor belt, as shown in Fig. \ref{fig:Transport_comparison}(b). The initial storage efficiency is the same as that in Fig. \ref{fig:Transport_inside_fiber}(b). The storage lifetime, however, is reduced by about 30\%, predominantly due to the weaker light-matter interaction when the atomic ensemble is transported away from the fiber tip, which results in a smaller retrival efficiency at longer transport distance. These two sets of data clearly show that the impact of experimental imperfections at the fiber tip is moderate. For instance, the stray light scattered on the fiber tip and the excitation of high-order modes can create strongly modulated potentials and heat the atoms during transport, causing additional atom loss or decoherence. Finally, we demonstrate the storage of light in an atomic medium which moves with constant speed during the storage process inside the HC-PCF, as shown in Fig. \ref{fig:Transport_comparison}(c). Here, we keep the conveyor belt running at a speed of $0.496~ \textrm{m/s}$ while storing and retrieving the light pulses.  This speed leads to a Doppler shift of $ -0.6~\textrm{MHz}$ for the probe and the control beams, which is in the same order as the linewidth of the employed lasers. In this case, we observe an increase of the storage lifetime by 30\% compared to the case of storing into a stationary ensemble [see Fig. \ref{fig:Transport_inside_fiber}(b)]. The initial storage efficiency, however, remains almost unchanged. From this, we conclude that the non-adiabatic abrupt acceleration and deceleration process induces some decoherence while the influence of the actual transport can be neglected. At the lifetime of $2.6~\textrm{ms}$, a storage efficiency of 3.6\% is obtained with transport distance of 1.26 mm, which is comparable to the size of the cloud itself.
Overall, we observe the high robustness of the storage efficiency and lifetime with respect to the parameters and details of the transport process. 

The experimental data show that light stored in the form of collective excitation can be actively transported over macroscopic distances in a highly controlled way. Currently, the transport distance is limited to several millimeters due to the short storage lifetime compared to the time necessary to transport the atomic medium. 
The storage time, in turn, is limited by the high heating and scattering rate caused by the broadband background spectrum of the tapered amplifiers used for boosting the power of the lattice beams \cite{Supplemental_material}. The scattering could be greatly suppressed by using another wavelength that is further away from the transition lines or avoiding the use of TAs. Significantly longer storage lifetime can be expected \cite{Dudin_2013}.
As the OD is a fundamental parameter that determines the storage efficiency \cite{Gorshkov_2007_Free}, increasing the atom number or using an HC-PCF with a smaller core size could increase the storage efficiency. The latter is also favorable for reducing the impact of micro-lensing effect \cite{Sulzbach2019}. Further optimization for the storage process itself is also possible by employing pulse shaping techniques \cite{Gorshkov_2007, Novikova_2007} or exploiting other atomic transition lines \cite{Hsiao_2018}.

Nevertheless, we demonstrate high robustness of the light storage against the dynamic acceleration and deceleration of the atomic ensemble. Although our experiments are done with coherent weak probe pulses, it can be extended to the regime of "true" quantum memory \cite{sprague2014broadband, Peters2020}. Evaluating the maximum allowed acceleration or transport speed might shed experimental light on the question of how fast stored quantum information can be transported. The work can also be extended to addressing several separated atomic ensembles sequentially or in parallel \cite{Roberts2014}. The control and probe beams need not necessarily be parallel to the fiber and flexible configurations are possible by using micromachined waveguides \cite{Elisa_2020}. This, for instance, offers the possibility of realizing a quantum race track memory for light or scalable quantum registers.

\begin{acknowledgments}
    We thank M. Noaman, M. Langbecker and D. Hu for earlier contributions to the experiment. We also thank Chris Perrella and Ben Sparkes for valuabe discussios.
    The work has been supported by the DFG SPP 1929 GiRyd and the UA-DAAD PPP Australia. W.L. acknowledges financial support by the Chinese Scholarship Council (CSC) and the Johannes Gutenberg-Universität Mainz. 
\end{acknowledgments}

\section{SUPPLEMENTARY MATERIAL}
\section{I: SAMPLE PREPARATION}
The 3-D MOT sketched in Fig. 1(a) of the main text is loaded from a 2-D MOT for 3 s. After sub-Doppler cooling, the atomic molasses is cooled down to $25 ~\mu \textrm{K}$  and contains $1.2\times10^7$ atoms with a size of about 1 mm ($1/e$ diameter). The optical lattice beams are switched on during the loading of the MOT. After transfer to the lattice, we wait 40 ms for the untrapped atoms to leave the interaction region. Afterwards, the atoms are pumped to state $\ket{g}$ by a $1~\textrm{ms}$ long pulse of the control beam. For optimal storage of the probe pulse, the atoms are further pumped to the $\ket{5S_{1/2},~F=1,~m=0}$ Zeeman state by sending another probe pulse of 0.5 ms duration but with its frequency turned close to resonance of the transition $\ket{5S_{1/2},~F=1} \rightarrow \ket{5P_{3/2},~ F'=0}$.

\section{II: PROBE AND CONTROL BEAMS}
To ensure phase stability, both the control and probe beams are derived from the same laser (Toptica, DL pro), which is frequency stabilized by Doppler-free spectroscopy and has a linewidth on the order of $100~ \textrm{kHz}$. As the two beams need to have a frequency difference of $6.834~\textrm{GHz}$, we pass part of the beam through a fiber-coupled Electro-Optic Modulator (EOM, iXblue NIR-MPX800-LN-05) driven at $6.5~\textrm{GHz}$ and isolate the first-order side-band. This is achieved by suppressing the unwanted frequencies by $30~\textrm{dB}$ with a temperature-stabilized single-lens etalon \cite{palittapongarnpim2012note}. For further control of the frequency and shaping of the pulses, the original and the frequency off-set beams are independently frequency shifted by double-pass AOMs. The probe and control beams are orthogonally polarized, spatially overlapped on a polarizing beam splitter and coupled to the fast and slow axes of a polarization-maintaining fiber, respectively.

We obtain a coupling efficiency into the HC-PCF of above 90\% for the optical lattice beams and 88\% for the probe and control beams. After passing through the HC-PCF, the probe and control beams are separated from the strong lattice beams by a dichroic-mirror and an optical band-pass filter (Thorlabs, FBH780-10). Probe and control beams are further separated with a polarizing beam splitter. The probe beam then passes through another temperature-controlled single-lens Etalon which further suppresses the control beam by 30 dB. 
In the end, we measure a total suppression ratio of $-47~\textrm{dB}$ of the control beam. The total transmission of the probe beam is $\approx 28\%$ when starting from the coupling into the HC-PCF to the power obtained at the fiber-coupled photomultiplier tube.

\section{III: LATTICE BEAMS}
The optical lattice beams are derived from a Ti: Sapphire laser working at $810 ~\textrm{nm}$ whose beam is split  and seeds two TAs. Each of the TA has an output power of about $2 ~\textrm{W}$. These TAs emit a broad background spectrum due to amplified spontaneous emission (ASE). A series of high pass filters with a cut-off wavelength of $800~\textrm{nm}$ (Thorlabs, FELH0800) are introduced into the lattice beams to partly suppress the ASE. The frequency detuning between the two lattice beams is controlled by two single-pass acousto-optical modulators (AOMs, MT80-A1.5-IR), driven by a programmable arbitrary waveform generator (FlexDDS from WieserLabs). The laser beams are coupled into two high-power polarization-maintaining fibers and delivered to the experimental chamber with a power of  $\approx 500~ \textrm{mW}$ each beam.

\section{IV: TRANSPORT OF ATOMS}
The driving frequencies of the AOMs are initially $80~\textrm{MHz}$. A moving lattice is obtained by ramping up one of the driving frequencies. The transport speed $\textit{v}$ can be determined from the  frequency detuning $\Delta \nu$ and wavelength of the lattice $\lambda$ by $\textit{v}=\Delta \nu \lambda /2$. Typically, we ramp up the frequency detuning linearly from 0 to $1200~ \textrm{kHz}$ in $1~ \textrm{ms}$, hold it constant for $14~ \textrm{ms}$ and ramp down the detuning for another $1~ \textrm{ms}$ to decelerate the atoms again [compare Fig. 1(c) in the main text]. This way, the atoms are transported $1~ \textrm{mm}$ into the fiber with a max acceleration of $486~\textrm{m/s}^2$ and a maximum speed of $0.486~\textrm{m/s}$. During the transport from the MOT position to the fiber tip, the power of the optical lattice beams is ramped down by $80\%$ to partly compensate the increasing trap depth due to the focusing of the beam towards the fiber tip. Compared to the case without the amplitude ramp, the amplitude-ramping process decreases the temperature of the atomic ensemble by half at the expense of losing about 50\% more of the atoms. In the latter experiments of transport of stored light, the dipole potential always stays constant.
The transport of atoms outside the HC-PCF can be visually confirmed by taking absorption images of the atomic cloud using an imaging beam propagating orthogonally with the lattice beams. However, we can not take absorption images of the atomic cloud after it is transported inside the fiber. To validate the transport of the cloud inside the HC-PCF, we measure the frequency detuning of the two optical lattice beams by beating the two driving frequencies of the AOMs and monitor the beat frequency with an oscilloscope. Details of the transport could be found in \cite{Langbecker2018}.

%

\end{document}